\documentclass[prb,aps,twocolumn,showpacs,amsmath,amssymb]{revtex4}
\usepackage{graphicx}% Include figure files
\usepackage{dcolumn}% Align table columns on decimal point
\usepackage{bm}% bold math

\begin{document}
\title{Pressure evolution of low-temperature crystal structure and bonding of 37 K $T_c$ FeSe superconductor}
\author{S. Margadonna$^1$}
 \email{serena.margadonna@ed.ac.uk, k.prassides@durham.ac.uk}
\author{Y. Takabayashi$^2$}
\author{Y. Ohishi$^3$}
\author{Y. Mizuguchi$^{4,5,6}$}
\author{Y. Takano$^{4,5,6}$}
\author{T. Kagayama$^7$}
\author{T. Nakagawa$^3$}
\author{M. Takata$^3$}
\author{K. Prassides$^2$,$^*$}
\affiliation{\\
$^1$School of Chemistry, University of Edinburgh, Edinburgh EH9 3JJ, UK
\\
$^2$Department of Chemistry, University of Durham, Durham DH1 3LE, UK
\\
$^3$Japan Synchrotron Radiation Research Institute, SPring-8, Hyogo 679-5198, Japan
\\
$^4$National Institute for Materials Science, 1-2-1 Sengen, Tsukuba 305-0047, Japan
\\
$^5$JST, TRIP, 1-2-1 Sengen, Tsukuba 305-0047, Japan
\\
$^6$University of Tsukuba, 1-1-1 Tennodai, Tsukuba 305-0001, Japan
\\
$^7$Center for Quantum Science and Technology under Extreme Conditions,
Osaka University, Osaka 560-8531, Japan
}

%\date{\today}

\begin{abstract}

FeSe with the PbO structure is a key member of the family of new high-$T_c$ iron pnictide and chalcogenide superconductors, as while it 
possesses the basic layered structural motif of edge-sharing distorted FeSe$_4$ tetrahedra, it lacks interleaved ion spacers or charge-reservoir layers.
We find that application of hydrostatic pressure first rapidly increases $T_c$ which attains a broad maximum of 37 K at $\sim$7 GPa (this is one of the highest 
$T_c$ ever reported for a binary solid) before decreasing to 6 K upon further compression to $\sim$14 GPa. 
Complementary synchrotron X-ray diffraction at 16 K was used to measure the low-temperature isothermal compressibility of $\alpha$-FeSe, revealing 
an extremely soft solid with a bulk modulus, $K_0$ = 30.7(1.1) GPa and strong bonding anisotropy between inter- and intra-layer directions that
transforms to the more densely packed $\beta$-polymorph above $\sim$9 GPa. 
The non-monotonic $T_c$($P$) behavior of FeSe coincides with drastic anomalies in the pressure evolution of the interlayer spacing, 
pointing to the key role of this structural feature in modulating the electronic properties.

\end{abstract}


\maketitle
\newpage

%\section{Introduction}
The $\alpha$-polymorph of the simple binary FeSe phase has recently emerged as a superconductor with an ambient $P$
$T_c$ of $\sim$8-13 K.\cite{hsu,mizuguchi} 
Its structure comprises stacks of edge-sharing FeSe$_4$ tetrahedra with a packing motif essentially identical to
that of the FeAs layers in the families of the FeAs-based high-$T_c$ superconductors\cite{Kamihara,Chenxh,Rotter,pitcher,tapp} 
but lacking any interleaved ion spacers or insulating layers.
The structural analogy is reinforced by the observation that below 70 K
the high-temperature crystal structure becomes metrically orthorhombic (space group $Cmma$), 
\cite{ChemComm2} displaying an identical distortion of the FeSe layers to that observed in the iron oxyarsenide family.
\cite{Cruz,Nomura} Theoretical calculations also find a very similar 2D electronic structure to that of the FeAs-
based superconductors with cylindrical electron sections at the zone corner and cylindrical hole surface sections.\cite{subedi}
Moreover, superconductivity in FeSe is very sensitive to defects and disorder and occurs
over a limited range of FeSe$_{1-\delta}$ non-stoichiometry.\cite{McQueen1} 

The effect of applied pressure on $T_c$
provides crucial information in differentiating between competing models of superconductivity and in
the FeSe binary, $T_c$ is initially extremely sensitive to $P$ and rises rapidly to 27 K at 1.48 GPa.\cite{mizuguchi}
At the same time, antiferromagnetic spin fluctuations present above $T_c$ are strongly enhanced by pressure.\cite{imai}
In the FeAs-based superconductors, the response of $T_c$ to pressurization 
is complex and sensitively depends on the composition of the materials and their doping level. 
Both positive and negative initial pressure coefficients,
d$T_c$/d$P$ have been measured. Typically for the REFeAsO$_{1-x}$F$_x$ families, d$T_c$/d$P$ is positive at low doping 
levels and switches over to a negative value as $x$ increases.\cite{takahashi,JACS,lorenz,takeshita,yi} 
Moreover, for systems where the initial d$T_c$/d$P$ is positive, there is a critical value of $P$ 
above which the trend is reversed and $T_c$ then decreases upon further pressurization.\cite{takahashi,alireza} 

Despite the importance of the evolution of $T_c$ with $P$ in understanding the
superconducting properties of the Fe-based materials, there is currently little information on the detailed
pressure dependence of their structural properties.\cite{garbarino,zhaoj,mito} Here we report on
the conducting properties of FeSe as a function of $P$ up to $\sim$14 GPa; we find a very high initial pressure coefficient
with $T_c$ maximized at 37 K at $\sim$7 GPa. At higher $P$, d$T_c$/d$P$ becomes negative and 
$T_c$ is reduced to 6 K at $\sim$14 GPa. 
We also study the precise pressure evolution to 12.8 GPa
of the FeSe structure at 16 K ({\it i.e.} well within the superconducting regime) 
by synchrotron X-ray powder diffraction. We find that the orthorhombic $\alpha$-polymorph survives to $\sim$9 GPa
whereupon the structure changes to that of the non-superconducting hexagonal $\beta$-phase. $T_c$ initially
increases with $P$ despite the decreasing SeFeSe thickness and the increasing distortion of the FeSe$_4$ units while the subsequent decrease 
clearly correlates with changes in the FeSe interlayer spacing.

%\section{Experimental details}

The FeSe sample used in this work was prepared, as reported elsewhere.\cite{mizuguchi}
Electrical resistivity measurements under high pressures were performed 
by a standard four-probe technique using a diamond anvil cell (DAC).
Powdered NaCl was used as pressure-transmitting medium. Pressure was applied at room temperature
and measured by the ruby fluorescence method. The measured values at low temperature are used
to discuss the pressure dependence of $T_c$.
The high-pressure synchrotron X-ray diffraction experiments at 16 K were performed at beamline BL10XU, SPring-8. 
The powder sample was loaded in a membrane DAC, which was placed inside a closed-cycle helium refrigerator. 
Daphne{\texttrademark} oil was used as a pressure medium. 
The applied pressure measured with the ruby fluorescence method was increased at 16 K without 
dismounting the cell from the cryostat. 
The diffraction patterns ($\lambda$ = 0.41118 \AA) were collected using a flat image plate detector 
up to 12.8 GPa. Data analysis of the diffraction profiles was performed with the GSAS suite of Rietveld programs.

%\section{Results}

\begin{figure}
\includegraphics[width=9cm]{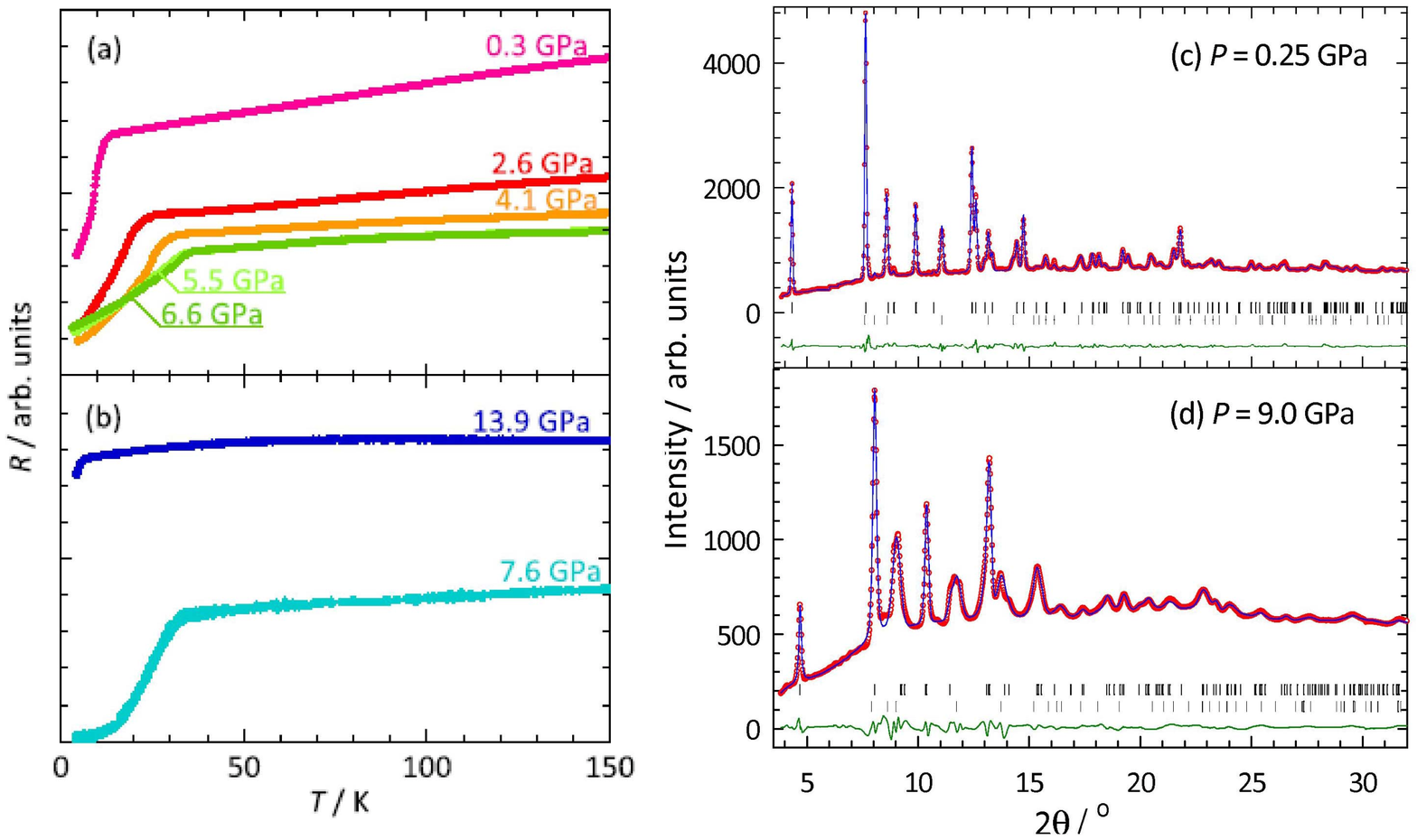}
\caption{(Color online) (a,b) Resistance against temperature for FeSe at several applied
pressures. (c), (d) Final observed (red circles) and calculated (blue solid line) synchrotron X-ray
($\lambda$ = 0.41118 \AA) powder diffraction profiles at 16 K for the FeSe sample at 
(c) 0.25 GPa and (d) 9.0 GPa. The lower green solid lines show the difference profiles and 
the tick marks show the reflection positions of the $\alpha$ (upper) and
$\beta$ (lower) phases, respectively.
\ }
\end{figure}

Fig. 1a shows the temperature dependence of the resistance of the FeSe sample at various pressures. 
Superconductivity was observed below 13.4 K with a fairly sharp onset. $T_c$ is found to increase initially rapidly
upon application of pressure with an accompanying increase in width and reaches a broad maximum of 37 K at 6.6 GPa $-$
for binary solids, this high-$T_c$ is only surpassed by Cs$_3$C$_{60}$ (38 K)\cite{alexey} and MgB$_2$ (39 K).\cite{MgB2}
In a second experiment, $P$ was increased first directly to 7.6 GPa and then to 13.9 GPa where the onset $T_c$ is smaller at 6 K (Fig. 1b).
Pressure release to near ambient and subsequent pressurization to 9.7 GPa resulted in smaller values of $T_c$ $-$
this may be related to the irreversibility of the $\alpha$$\rightarrow$$\beta$ structural transformation ({\it vide infra}) (Fig. 2d).

Inspection of the diffraction profile\cite{epaps} at 16 K and 0.25 GPa (Fig. 1c) readily
reveals the orthorhombic (O) unit cell (space group $Cmma$)
established for $\alpha$-FeSe below 70 K at ambient $P$.\cite{ChemComm2}
Additional peaks are also evident and these can be accounted for by
the presence of a minority hexagonal NiAs-type $\beta$-FeSe phase (32.4(1)\% fraction).
The diffraction datasets collected with increasing $P$ show no structural
changes for $\alpha$-FeSe between 0.25 and 7.5 GPa with the $\alpha$/$\beta$$-$phase assemblage 
remaining unaltered.\cite{epaps} The same structural model was therefore employed in the Rietveld refinements in this pressure range,
revealing a monotonic decrease in the lattice constants and unit cell volume with increasing $P$ (Fig. 2a,b).  
However, the response of the lattice metrics
to $P$ is strongly anisotropic with the interlayer spacing showing a significantly larger contraction than the
intralayer dimensions. As $P$ increases, 2$c$/($a$+$b$) smoothly decreases 
until at $\sim$4 GPa, it approaches 1. 
Further increase in $P$ leads to an even higher compression of the FeSe interlayer spacing (Fig. 2c) with
[2$c$/($a$+$b$)] = 0.986(2) at 7.5 GPa.
However, as the sample is pressed to 9.0 GPa (Fig. 1d), the $\alpha$-phase fraction begins to decrease and then it sharply collapses with an almost complete
$\alpha$$\rightarrow$$\beta$ transformation taking place.\cite{epaps} At the same time, while the basal plane 
lattice constants continue to contract, $c$ begins to increase, resulting in saturation of the volume
response to pressure and an increasing 2$c$/($a$+$b$) ratio (Fig. 2a-c). 
Gradual depressurization to $\sim$2 GPa does not lead to recovery of the orthorhombic 
$\alpha$ polymorph, implying that the $\alpha$$\rightarrow$$\beta$ phase transformation is irreversible 
at these low temperatures. Recovery of $\alpha$-FeSe (53\% fraction) at this $P$ necessitated heating of the sample inside the 
MDAC to 300 K.\cite{NiAs}

\begin{figure}
\includegraphics[width=9cm]{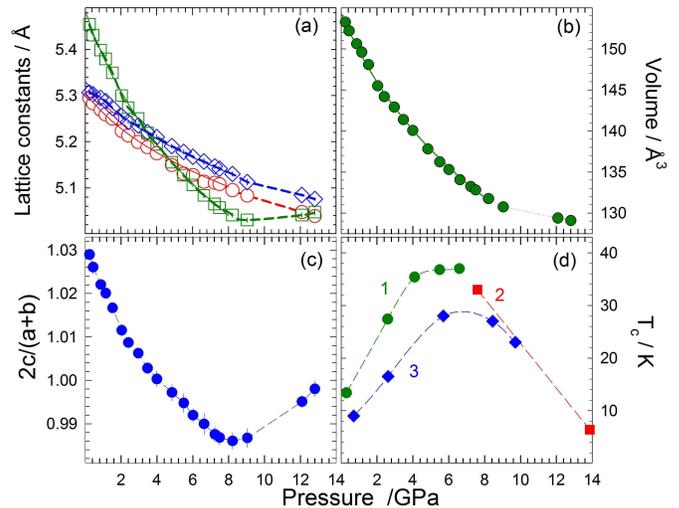}
\caption{(Color online) Pressure dependence of: (a) the orthorhombic lattice constants $a$ (circles), $b$ (diamonds), 
and $c$ (squares),
(b) the unit cell volume, $V$, (c) the [2$c$/($a$+$b$)] ratio, and (d) the superconducting transition temperature, $T_c$ of $\alpha$-FeSe.
The solid lines through the data points to 7.5 GPa in (a) and (b) are least-squares fits 
to the Birch-Murnaghan equation-of-state, while the dotted lines
are guides to the eye. 
\ }
\end{figure}

Fig. 2b shows the pressure evolution 
of the unit cell volume of $\alpha$-FeSe together with a least-squares 
fit of its 16 K equation-of-state (EOS) to the semi-empirical third-order Birch-Murnaghan equation.\cite{EOS}
The fit results in values of
the atmospheric pressure isothermal bulk modulus, $K_0$= 30.7(1.1) GPa and 
its pressure derivative, $K_0^{'}$= 6.7(6) taking into account the $V$($P$) data to 7.5 GPa. The volume compressibility, 
$\kappa$ = dln$V$/d$P$= 0.033(1) GPa$^{-1}$ implies a soft highly compressible solid. 
A comparable value of the bulk modulus $\sim$33 GPa at 50 K has been proposed from neutron diffraction
measurements over a restricted 0-0.6 GPa range.\cite{millican}
The anisotropy in bonding of the $\alpha$-FeSe structure (Fig. 3d) 
is clearly evident in Fig. 2(a) which displays the variation of the orthorhombic  
lattice constants with $P$. $\alpha$-FeSe
is least compressible in the basal plane, in which the 
covalent Fe-Se bonds lie (dln$a$/d$P$= 0.029(2) GPa$^{-1}$, dln$b$/d$P$= 0.026(3) GPa$^{-1}$), 
while the interlayer compressibility, dln$c$/d$P$= 0.065(4) GPa$^{-1}$ is $\sim$2.5 times larger, 
implying very soft Se-Se interlayer interactions.

%\section{Discussion}
The immediate consequence of the structural simplicity of FeSe is that the $\sim$2.91 \AA-thick SeFeSe building blocks are 
in close proximity separated by each other by only $\sim$2.58 \AA\ (Fig. 3d).\cite{ChemComm2} This contrasts
sharply with the enhanced interlayer separation in the FeAs analogues (e.g. interslab separation
is $\sim$5.79 \AA\ in SmFeAsO which comprises interleaved SmO layers\cite{PRB} and $\sim$3.45 and 3.34 \AA\ in SrFe$_2$As$_2$ and
LiFeAs which possess AE$^{2+}$ and Li$^+$ spacers, respectively).\cite{Rotter,pitcher} As a result, the $c$ axis in FeSe is only
marginally larger (1.03 times) than the average basal plane dimensions. The electronic consequence of this structural size proximity should be
to render the electronic structure of FeSe more 3D in nature than those of any of the related FeAs superconductors.
Moreover, given the softness of the interlayer Se-Se interactions relative to the covalently bonded
SeFeSe slabs, application of pressure should have a profound influence on the structural and electronic
dimensionalities, allowing their tuning at interlayer contact separations inaccessible in other currently known Fe-based
superconductors. This is of key importance if we recall both the extreme sensitivity of $T_c$ to $P$ and its non-monotonic response at high $P$.
The distinct structural response to $P$ of $\alpha$-FeSe is clearly apparent 
by considering the compressibility and its anisotropy (Fig. 2a,b). Firstly, the low-$T$ bulk modulus, $K_0$ =
30.7(1.1) GPa is the
smallest measured thus far in any of these systems $-$ $K_0$ is 57.3(6) GPa for LiFeAs,\cite{mito} 
78(2) GPa for LaFeAsO$_{0.9}$F$_{0.1}$, and 102(2) GPa for NdFeAsO$_{0.88}$F$_{0.12}$.\cite{garbarino,zhaoj} In addition, 
the compressibility of $\alpha$-FeSe along the interlayer direction is the largest amongst these systems $-$ the
$c$ axis contracts by 7.3\% at 7.5 GPa $-$ while the basal plane compressibility (contraction by 3.3\% at 7.5 GPa) is 
comparable to that in LiFeAs but considerably larger than those of the iron oxyarsenides.

\begin{figure}
\includegraphics[width=9cm]{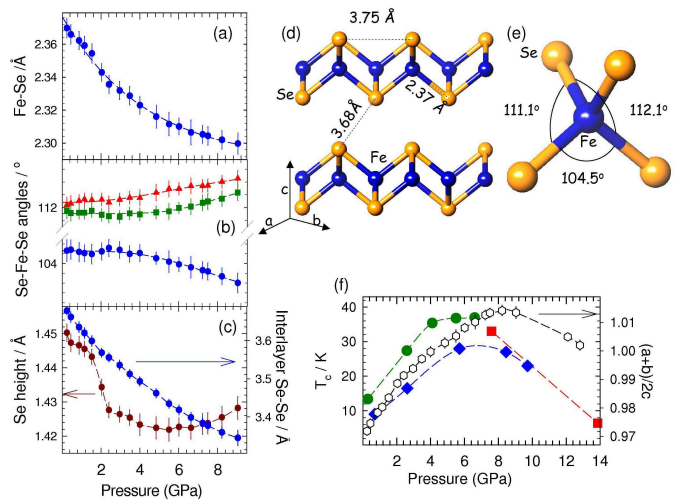}
\caption{(Color online) (a-c) Pressure dependence of: (a) the Fe-Se bond distances,
(b) the three distinct Se-Fe-Se bond angles, and (c) the inter-layer Se-Se contacts and the height of the Se atoms above the Fe plane 
for the orthorhombic crystal structure of $\alpha$-FeSe at 16 K. 
(d) Schematic diagram of the crystal structure of $\alpha$-FeSe with selected distances at 16 K and 0.25 GPa. 
(e) Geometry of the FeSe$_4$ units and definition of the three Se-Fe-Se bond angles together with their values at
16 K and 0.25 GPa. (f) Pressure dependence of the [($a$+$b$)/2$c$] ratio at 16 K and the superconducting $T_c$ of
$\alpha$-FeSe. The lines through the points are guides to the eye.
\ }
\end{figure}

This sets the scene to discuss the pressure response of $T_c$ in $\alpha$-FeSe and its relationship to the structural
evolution with $P$. Fig. 3a,3c 
show the pressure dependence at 16 K of the Fe-Se and Se-Se interatomic distances. The Fe-Se bonds contract smoothly to $\sim$4 GPa
but remain essentially unchanged with further increase in $P$ resulting in an overall decrease of 2.3\% at 9.0 GPa. 
Similarly the intralayer Se-Se distances decrease monotonically with a somewhat
larger contraction of 3.7\%. However, the pressure response of the interlayer Se-Se contacts is much steeper with a 9.8\% decrease
reflecting the very large interlayer compressibility $-$ the SeFeSe slabs approach each other rapidly 
up to 7.5 GPa but then their contact distance appears to saturate. As a result, there is a correlation between
the pressure evolution of the unit cell metrics and the superconducting $T_c$ (Fig. 3f). At low pressures, the rapid increase
in ($a$+$b$)/2$c$
is mirrored by the rapidly increasing $T_c$ ($\alpha$-FeSe has the smallest bulk modulus 
and highest pressure coefficient of $T_c$ amongst the Fe-based superconductors). Remarkably the pressure range at which the maximum
$T_c$ is found coincides with the onset of the structural anomalies. While the basal plane lattice constants
continue to contract smoothly, the interlayer spacing begins to expand slightly leading to a decreasing ($a$+$b$)/2$c$ ratio 
as $P$ increases above 7.5 GPa. Thus it is tempting to ascribe the non-monotonic $T_c$($P$) bahavior
to the competition between the effects of the interlayer SeSe interactions (tuning the doping level) and the intralayer FeSe bonding and
suggests that higher $T_c$s are associated with smaller interlayer spacings (increased dimensionality). 

Fig. 3b,c shows the pressure dependence at 16 K of the thickness of the SeFeSe slabs and the crystallographic bond angles of the FeSe$_4$ tetrahedra (Fig. 3e). 
It has been argued for the iron (oxy)arsenides that 
the geometry of the edge-sharing FeAs$_4$ tetrahedral units sensitively controls 
the width of the electronic conduction band and therefore the magnitudes of the slab thickness and the As-Fe-As angles are important parameters in tuning
the electronic properties of these systems and determining $T_c$.\cite{Zhao,McQueen,lee} Empirically 
$T_c$ appears to be maximal when the FeAs$_4$ units are close to regular with As-Fe-As angles 
of 109.47$^{\circ}$.\cite{lee} Along these lines, the observation of negative pressure coefficients of $T_c$ has been rationalized 
in terms of increased tetrahedral distortion away from regular shape.\cite{mito,zhaoj,kumai} 
$\alpha$-FeSe has an identical distortion mode of the FeSe$_4$ tetrahedra with a sizable distortion away
from regularity (7$^{\circ}$ angle difference, Fig. 3e). However, while the distortion also increases 
significantly with increasing $P$ and the SeFeSe slab thickness decreases (Fig. 3b,c), the pressure coefficient
of $T_c$ is positive in this range, implying that the empirical generalization which appears to hold well for the iron (oxy)arsenides is not
applicable in the present system $-$ apparently the dependence of the electronic structure on the interlayer
separation dominates in determining the superconducting properties of $\alpha$-FeSe.
 
Finally, we discuss the high-$P$ $\alpha$$\rightarrow$$\beta$ transformation. Firstly, 
pressurization leads to a rapid decrease in the Se-Se interlayer contacts in the $\alpha$-phase $-$ at
7.5 GPa, these approach a value of 3.3 \AA, which is significantly smaller than the sum of the Se$^{2-}$ ionic radii.
As a result, the sign of the compressibilty along $c$ changes and the interlayer spacing begins to increase upon further
increase in $P$.
In addition, the three-dimensional $\beta$-FeSe is more densely packed than the quasi-two-dimensional $\alpha$ polymorph. 
The large difference in packing density
is retained at high $P$ and at 9.0 GPa, the $\alpha$$\rightarrow$$\beta$ transition is accompanied by a $\sim$14.6\% decrease in volume
as the Fe-Se coordination changes from tetrahedral to the more densely packed octahedral one.
At the same time, the transition results in an increase of the Fe-Fe and Fe-Se distances by 6.4\% and 4.0\%, respectively. 
These observations point toward a scenario where the $\alpha$-phase cannot sustain pressures above 9 GPa 
as the interlayer separation is now so small that is energetically favorable to form the more efficiently 
packed $\beta$-phase with a release of the bonding distances.

%\section{Conclusion}

In conclusion, we have found that the superconductivity onset in $\alpha$-FeSe attains a broad maximum of 37 K at $\sim$7 GPa.
The orthorhombically distorted superconducting phase is extremely soft and the pressure response of the structure
reveals an intimate link between the SeFeSe interlayer separations and the superconducting properties.
Detailed band structure calculations should be able to shed 
light on the accompanying evolution of the doping level of 
the SeFeSe layers and decipher the relative importance between the increased dimensionality and the
FeSe slab geometry to superconductivity.

%\section{Acknowledgement}
We thank SPring-8 for access to the synchrotron X-ray facilities and
K. Kuroki for useful discussions.


\begin{thebibliography}{00}
\bibitem{NiAs}
$\beta$-FeSe undergoes a phase transition to a lower symmetry structure at 6.0 GPa. 
The observed Bragg reflections are well accounted for by a MnP-type orthorhombic structure (space group $Pbnm$)  
with lattice constants related to those of the NiAs-type hexagonal unit cell by 
$a_o$ = $\sqrt{3}a_h$, $b_o$ = $c_h$, and $c_o$ = $a_h$.
\end{thebibliography}
\end{document}